\documentclass[draft]{agujournal2019}
\usepackage{url} 
\usepackage{lineno}
\usepackage[finalnew]{trackchanges} 
\usepackage{soul}
\usepackage{multirow}
\draftfalse

\journalname{JGR: Planets}

\begin{document}

%
%


\title{Mapping the Brightness of Ganymede's Ultraviolet Aurora using
  Hubble Space Telescope Observations}

%
%




\authors{A. Marzok\affil{1}, S. Schlegel\affil{1}, J. Saur\affil{1}\thanks{}, L. Roth\affil{2}, D. Grodent\affil{3}, D.F. Strobel\affil{4}, K.D. Retherford\affil{5,6}}

\affiliation{1}{University of Cologne, Germany}
\affiliation{2}{KTH Royal Institute of Technology, Sweden}
\affiliation{3}{University of Li\`{e}ge, Belgium}
\affiliation{4}{Johns Hopkins University, USA}
\affiliation{5}{Southwest Research Institute, USA}
\affiliation{6}{University of Texas at San Antonio, USA}






\correspondingauthor{Joachim Saur}{jsaur@uni-koeln.de}




\begin{keypoints}
\item Brightness map of Ganymede's ultraviolet auroral emission 
  \add{ 
    has been constructed based on a
  } 
  large set of HST observations
\item Auroral ovals are structured in upstream and downstream 'crescents'
\item Brightness on sub-Jovian and anti-Jovian side is strongly reduced by a factor of 3 - 4 compared to upstream and downstream side
\end{keypoints}

%
%

%
%


\begin{abstract}
We analyze Hubble Space Telescope (HST) observations of Ganymede 
made with the Space
Telescope Imaging Spectrograph (STIS) between 1998 and 2017 to
generate a brightness map of
Ganymede's oxygen  emission at 1356 {\AA}. Our Mercator projected map demonstrates
that the brightness along Ganymede's northern and southern auroral
ovals strongly varies with longitude. 
To quantify this variation around Ganymede, we investigate the
brightness averaged over 36$^{\circ}$-wide longitude corridors centered around
the sub-Jovian (0$^{\circ}$ W), leading (90$^{\circ}$ W), anti-Jovian
(180$^{\circ}$ W), and
trailing (270$^{\circ}$ W) central longitudes.
In the northern hemisphere, the brightness of the auroral oval is 3.7 $\pm$
0.4 times lower in the sub-Jovian and anti-Jovian corridors compared
to the trailing and leading corridors.
The southern oval is overall brighter than the northern oval,  and 
only  2.5 $\pm$ 0.2 times fainter on the sub- and
anti-Jovian corridors compared to the trailing and leading corridors. 
This demonstrates
that Ganymede's auroral ovals are strongly structured in auroral
crescents on the leading side (plasma downstream side) and on the
trailing side (plasma upstream side). We also find that the
brightness is not symmetric with respect to the 270$^\circ$ meridian, 
but shifted by $\sim$20$^\circ$ towards the
Jovian-facing hemisphere.
Our map will be useful for subsequent
studies to understand the processes
\change{ 
  which  
}{  
  that
}  
generate the aurora in  Ganymede's 
non-rotationally driven, sub-Alfv\'{e}nic
magnetosphere.
\end{abstract}

\section*{Plain Language Summary}
Northern lights often illuminate the night sky in a shimmering green
or red tone at high geographic latitudes. This emission,
scientifically referred to as \textit{aurora}, is a result of
electrically charged particles that move along
\change{ 
  the planets'  
}{ 
  Earth's
} 
magnetic
field
\add{ 
  lines
} 
and interact with its atmosphere to produce auroral
emission. Apart from the Earth, multiple
\add{ 
  other
} 
planets in our solar system
also
\change{ 
  possess  
}{ 
  exhibit
} 
auroral emission. By characterizing the
brightness and structure of these lights, we are therefore able to
deduce insights about a planet's atmosphere, magnetic field and the
physical processes occurring along the field lines from afar.
In this work, we used observations from the Hubble Space Telescope to
analyze the auroral emission of Jupiter's largest moon Ganymede. We
combined multiple images of Ganymede to create the first complete map
that displays the auroral brightness.  Our
map revealed that the emission on Ganymede's
\add{ 
  auroral
} 
ovals varies strongly in
brightness with divisions into two distinct bright and faint regions.
They resemble two auroral crescents in the north and south respectively,
and demonstrate the uniqueness of Ganymede's aurora in comparison with
the auroral ovals of other planets in the solar system.

%
\section{Introduction}
\label{s:intro}
Being the only known moon in our solar system with an internal dynamo
magnetic field \cite[]{kive96,kive02}, Jupiter's largest satellite Ganymede
\change{ 
  possesses  
}{ 
  exhibits
} 
auroral emission structured by its magnetic field. 
The first hint of polar auroral emission
\change{ 
  around  
}{ 
  at
} 
Ganymede was found by \citeA{hall98} who used the Goddard
High Resolution Spectrograph on the Hubble Space Telescope (HST) to
observe Ganymede's trailing hemisphere in the FUV. The retrieved peaks in the spectrum
around 1304 {\AA} and 1356 {\AA} were interpreted as emission from
a tenuous oxygen atmosphere. 
The observed double-peak profile of the
1356 {\AA} emissions indicated that the emissions are spatially
confined to the moon's magnetic north and south poles, suggesting auroral
emissions \cite{hall98}. The species responsible for the emissions was determined
from the detected flux ratios of OI 1304 {\AA} and OI 1356 {\AA} to be
primarily molecular oxygen via dissociative electron-impact
excitation. \citeA{feld00a} first imaged the auroral emission 
with the Space Telescope Imaging Spectrograph (STIS)
\remove{installed} 
on the HST. The obtained images of the upstream hemisphere
depicted diffuse background emission  with localized bright
regions of 300 R at latitudes of approximately $\pm 40^{\circ} $. Evaluating Galileo
spacecraft data, \citeA{evia01} argued that the measured
population of thermal electrons $n_e \approx 5 - 20$ cm$^{-3}$ with a
temperature of 20 eV are not able to create even the diffuse
background emission and that existing supra-thermal electrons of 2
keV are too few with a density of only $n_e \approx 0.5 - 2$
cm$^{-3}$ to be responsible for the aurora as well. 
Therefore an additional process is required to accelerate
the electrons to sufficient energies that could produce the emission.

\begin{figure}[b!]
\noindent\includegraphics[width=0.8\textwidth]{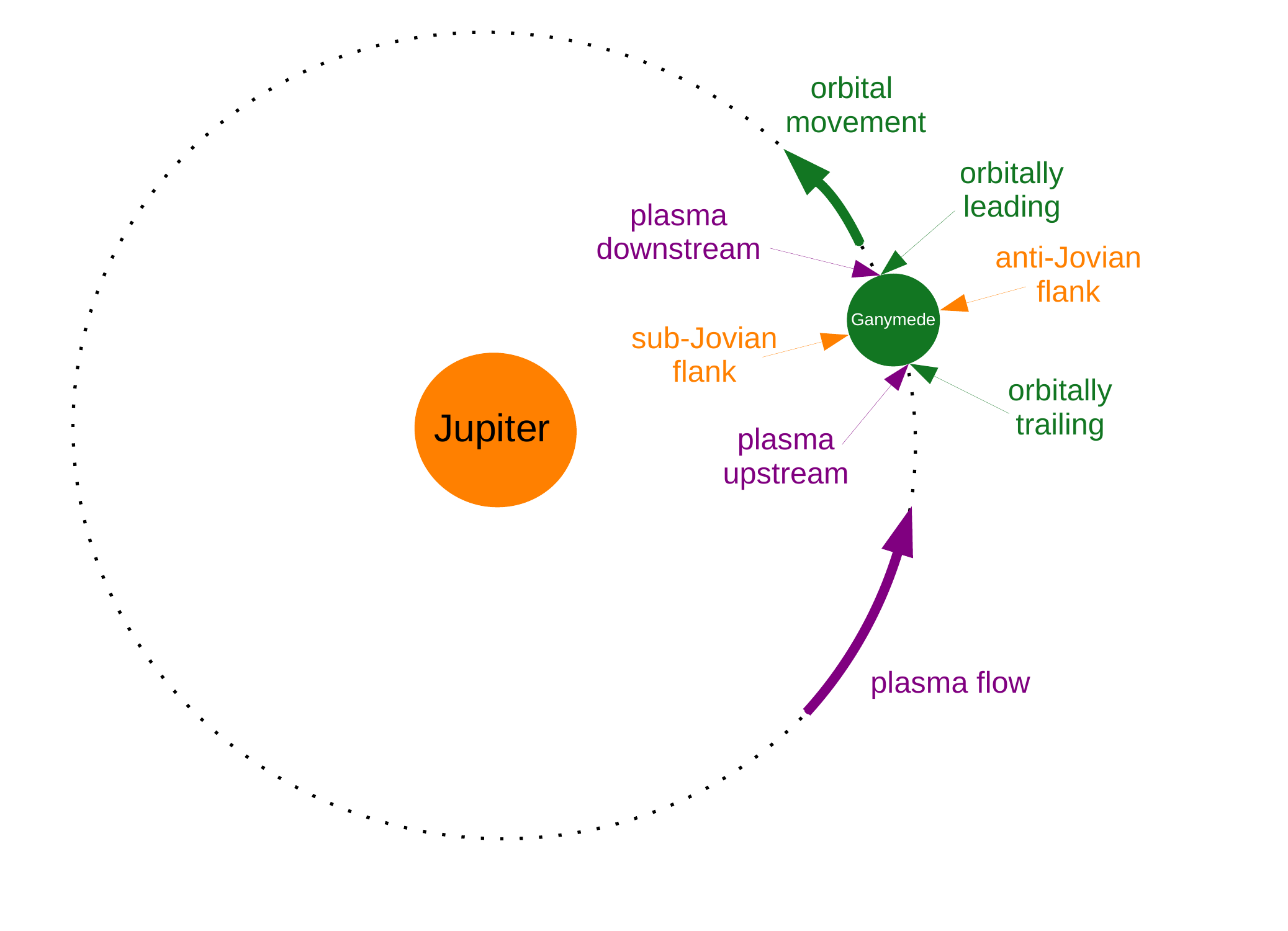}
\caption{Geometry of orbital direction, plasma flow and related
  terminology for the various hemispheres/sides of Ganymede. The
  plasma flow is faster than Ganymede's orbital velocity and
  therefore the trailing side is the upstream side of the plasma
flow.}
\label{f:geom}
\end{figure}

From collected HST observations of the downstream and upstream
hemispheres, \citeA{mcgr13} created a map of the location of
Ganymede's auroral bands at 1356 {\AA}. Their results showed that
the emission is correlated with Ganymede's plasma environment. The
magnetospheric plasma of Jupiter is approximately 
corotating with
\change{ 
  it's  
}{ 
  its
} 
magnetic
field at a synodic rotation period of 10.5 hours. As Ganymede is orbiting
Jupiter in a synchronized rotation period of only 7.2 days, the bulk
plasma flow therefore overtakes the moon on its orbit. On the orbitally
trailing hemisphere, where the plasma streams towards the moon, the
\add{ 
  auroral
} 
bright spots are mapped to latitudes of $40^{\circ}- 55^{\circ}$. On
the other hemisphere, i.e., the downstream hemisphere, the brightest auroral
emissions are found to be much closer to the equator near
latitudes of only $10^{\circ} - 30^{\circ}$
\cite[]{mcgr13,musa17}. In this work we use the terminology 'orbitally
leading side' which corresponds to the 'plasma downstream side' and 'orbitally trailing
side' which corresponds to the 'plasma upstream side'
\change{ 
  interchangeable  
}{ 
  interchangeably,
}  
depending on the physical context.
\add{ 
  These hemispheres are displayed in Figure
} 
\ref{f:geom}
\add{ 
  for  visual orientation.
  } 

The aurora was further characterized by \citeA{musa17}, who
also collectively analyzed STIS HST images from 1998, 2000, 2010 and
2011. Their findings include that the aurora changes position with
the moon's changing location in Jupiter's magnetosphere. 
When Ganymede is inside the Jovian current sheet, the upstream emission
is shifted by $+2.9^{\circ}$ towards higher latitudes and by $-4.1^{\circ}$ towards the
equator on the downstream hemisphere. 
The authors also found that
the total disk 
brightness is on average 1.4 times
\change{ 
  larger  
}{ 
  greater
} 
on the downstream side than on the
upstream side. When Ganymede is located
inside  the current sheet compared to outside the current sheet, 
the brightness on the 
downstream side increases by a factor of 1.5 and decreases by 0.8 on
the upstream side.
By averaging the temporal effects of the various
observed positions
\change{ 
  on  
}{ 
  of
} 
the aurora, \citeA{musa17} further characterized
Ganymede's internal magnetic field and derived a modified position for the
longitude of its dipole. The temporal effects were also
studied by \citeA{saur15}, who used the oscillation of the aurora to
demonstrate the presence of an ocean beneath Ganymede's icy crust from HST
observations. The oscillation arises from Jupiter's time-varying
magnetic field in the rest frame of Ganymede. Further details about
the hydrogen corona and oxygen atmosphere of Ganymede were published
recently in the works of \citeA{moly18} and
\citeA{alda17}. \citeA{alda17} used data from 4 STIS campaigns
between 1998 and 2014 to determine the abundance and variation of
atomic hydrogen around Ganymede by analyzing the detected
Lyman-$\alpha$ emissions. \citeA{moly18} used 
observations obtained with the Cosmic Origins Spectrograph
(COS) along with STIS data to characterize the variations
in the emission and the composition
of Ganymede's oxygen atmosphere on the leading and trailing sides from
measured intensities at 1304 {\AA} and 1356 {\AA}.
Very recently, \citeA{roth21}
\change{ 
  discovered  
}{ 
  found evidence of
} 
water vapor in Ganymede's
atmosphere and found that near the subsolar point sublimated water
vapor is more abundant than than molecular oxygen.

Various numerical simulations of Ganymede's magnetic field and plasma
environment  contribute to the understanding of its auroral
emission. \citeA{kopp02} applied
resistive magnetohydrodynamic (MHD) simulations to show that the
open-closed field boundary (OCFB) is changing with respect to the
varying magnetic environment around Ganymede. The OCFB marks the
separatrix
between those magnetic field lines of Ganymede that close
on the moon and those that are connected to Ganymede on one end and to
Jupiter on the other \cite{neub98}. Due to the magnetospheric plasma flow and the
associated magnetic stresses, the OCFB on the upstream side is
shifted to higher latitudes while it is dragged towards the equator on
the downstream side. \citeA{jia08} used single-fluid MHD
simulations to describe the interaction of Ganymede's magnetosphere
with the ambient magnetic field. Their findings indicate that the major
process for plasma and energy to enter the magnetosphere is via magnetic
reconnection that occurs on the down- and upstream sides, where
ambient and intrinsic field lines are nearly anti-
parallel. Reconnection primarily occurs at
the magnetopause on the upstream side and in a thin equatorial region
on the downstream side which extends several Ganymede radii away. 
The comparison between the observed location of Ganymede's peak
auroral emission by \citeA{mcgr13} and MHD
modeling of Ganymede's environment performed by \citeA{jia08} showed
that the locations of Ganymede's auroral ovals are well correlated
with the OCFB of Ganymede's magnetic field lines. 
\citeA{duli14}
also modeled Ganymede's plasma interaction with an MHD
model with a new description for the insulating boundary conditions on
Ganymede's icy surface. The resultant location of the OCFB for various
upstream conditions in \citeA{duli14} and \citeA{jia08} are very
similar as discussed in \citeA{saur15}. Additionally, 
3D multi-fluid MHD simulations  were applied 
\cite[]{paty04,paty06} or hybrid models \cite[]{fate16} 
were used to estimate neutral
sputtering rates on the surface \cite[]{liuz20}.
Further models focused on additional plasma effects of Ganymede's
magnetosphere such as Hall MHD \cite{dore15}. \citeA{toth16,
zhou19, zhou20} used embedded particle-cell and MHD models to better
understand reconnection at Ganymede and the resultant energetic
particle fluxes. For all these models, the structure and
brightness of Ganymede's auroral belts, the subject of this work,
are key observational constraints (next to Galileo in-situ measurements) to
understand the physics of Ganymede's sub-Alfv\'{e}nic
mini-magnetosphere.

While a location map of the aurora was created by \citeA{mcgr13},
and the time-variable aspects of Ganymede's aurora, as well as the
local emission morphology was studied by \citeA{saur15} and
\citeA{musa17}, in this work we create a first complete global
Mercator map of Ganymede's auroral brightness at
OI 1356 {\AA}. Here we also use previously unpublished
HST observations from 2017 to explicitly focus on the emission
structure at the sub- and anti-Jovian flanks. We use the emissions at
OI 1356 {\AA} because it provides the largest signal-to-noise ratio
compared to OI 1304 {\AA} \cite[]{musa17}. The brightness
structure is analyzed with special regard to the continuity of both
ovals. Our map will  serve as a diagnostic tool for future studies
of magnetospheric and auroral processes around Ganymede.

\section{Observations and Data Processing}
In this section we describe the HST/STIS datasets which were used in
our study. We also describe how
we map auroral emission from Ganymede's disk onto a Mercator map.

\subsection{Overview of the Observations}
\begin{figure}[b!]
\noindent\includegraphics[width=0.45\textwidth]{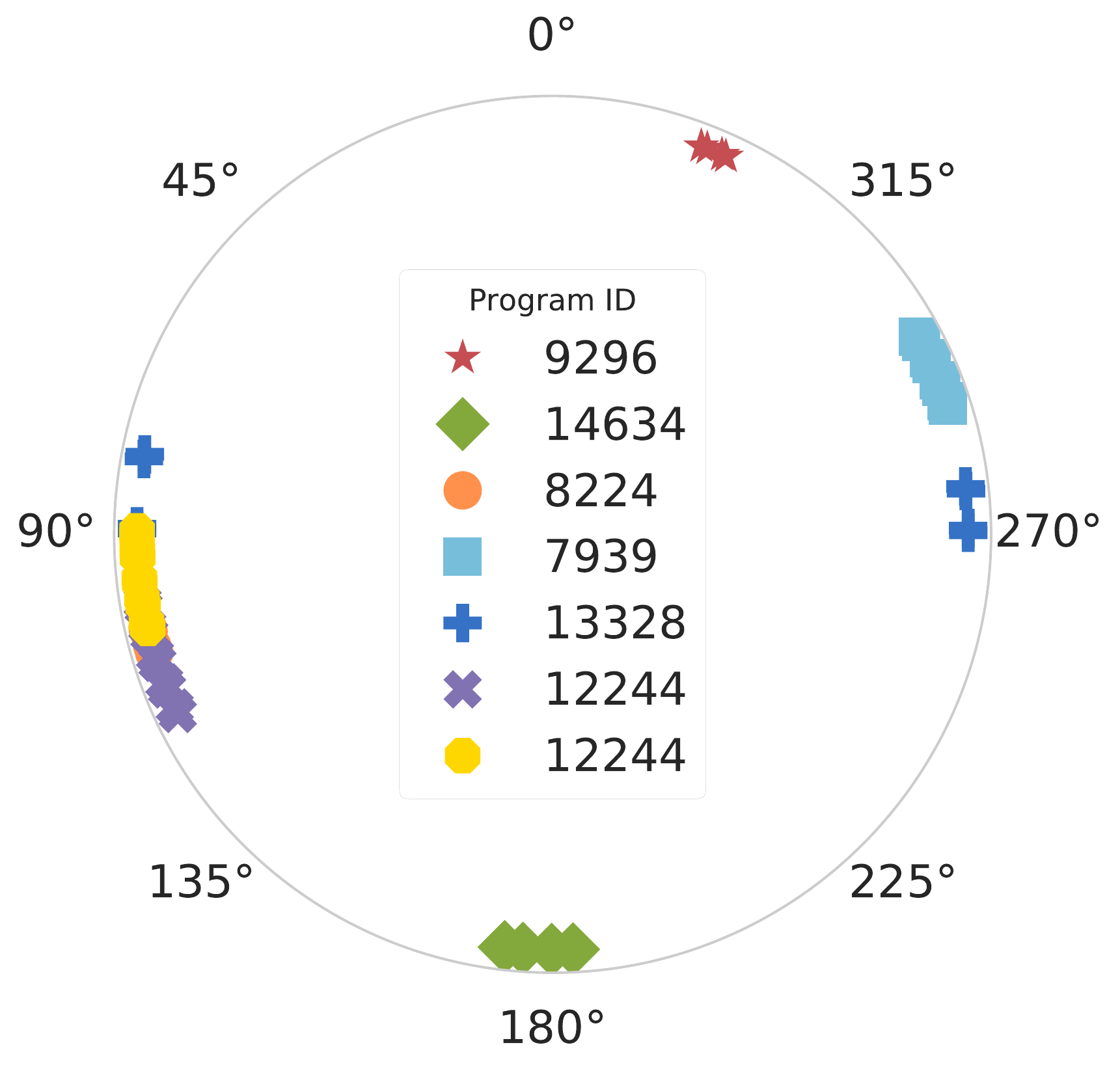}
\caption{Overview of the elongation for all available exposures of
  each program, listed by their Program ID.
  \add{ 
    Program 12244 consists out
    of two visits, which are displayed separately.
  } 
}
\label{fig:elong_ovr}
\end{figure}
\begin{figure}[t!]
\noindent\includegraphics[width=\textwidth]{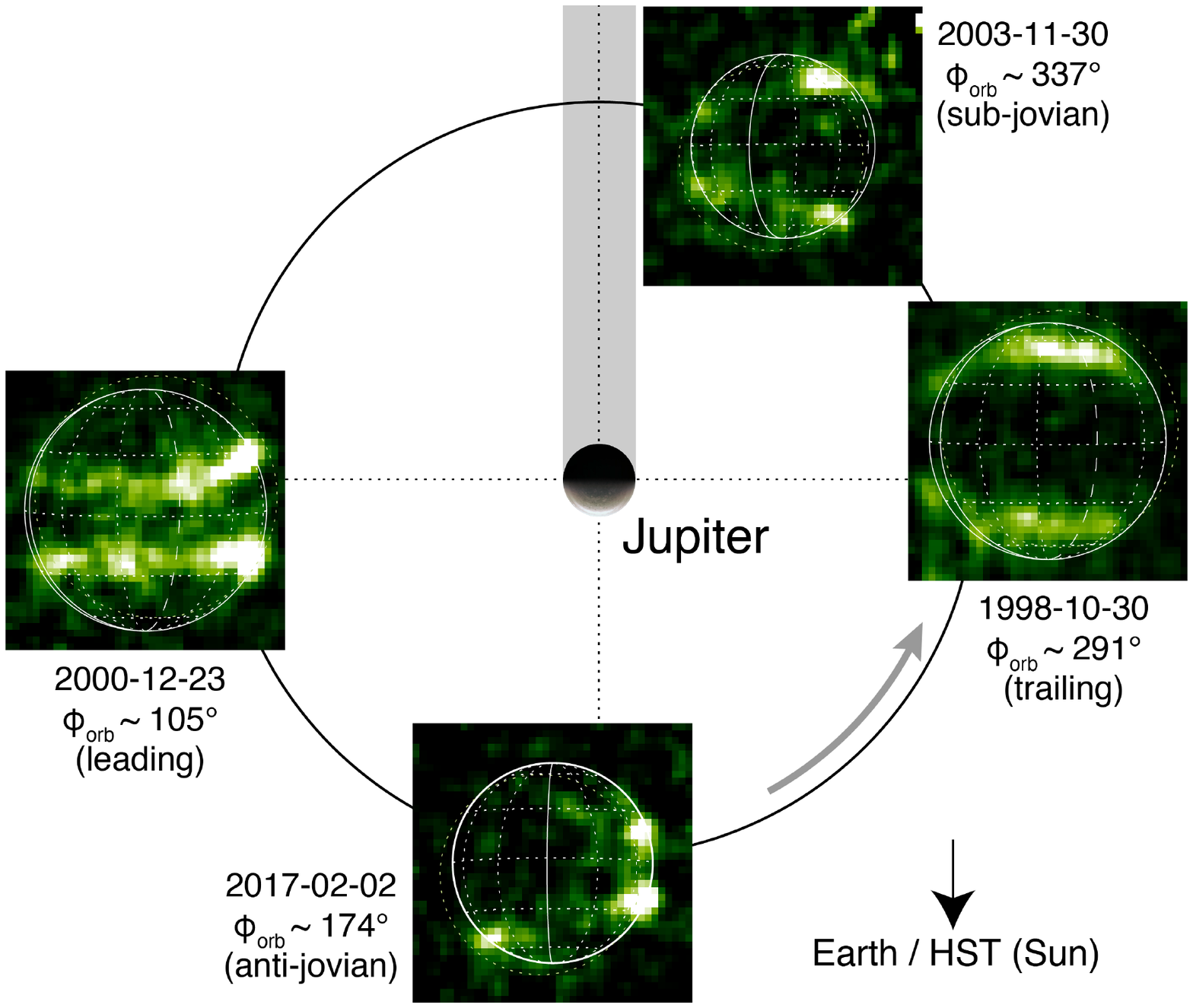}
\caption{ Selected observations of Ganymede's auroral emission at OI
  1356 {\AA} showing Ganymede's leading, trailing, sub-Jovian and
  anti-Jovian side. The auroral ovals are closer to the equator on the
  leading side compared to the trailing side, but appear continuous in
  both cases. In contrast, on the sub-Jovian and anti-Jovian side the
  aurora appears to be interrupted near 0$^{\circ}$ and 180$^{\circ}$
  (meridians as solid white lines).
  \add{ 
    The individual images of this
    Figure have been generated from the original HST data. The
    observations at the anti-jovian geometry have
    not been published before,
  while observations at other orbital longitudes have already been displayed
  in the work discussed in Section
} 
\ref{s:intro}. 
}
\label{fig:images}
\end{figure}

\begin{table}
\caption{Available programs where Ganymede was observed with
  HST/STIS. $\theta_{mag}$ is Ganymede's 
magnetic latitude in Jupiter's magnetosphere, 
$\lambda_{obs}$ the sub-observer planetocentric latitude of HST and $\phi_{orb}$
the elongation of Ganymede around Jupiter as in Figure
\ref{fig:elong_ovr}. Orbit refers to the number within a visit.
\add{ 
  Program 12244 consists of two visits with 5 orbits each taken approximately one
  year apart.
} 
}
\label{tab:exp_ovr}
\begin{tabular}{l c c c r c c c c c}
\hline
\hline
ID & Date  & Orbit & Exposures ID & $\lambda_{III}$&
                                                                 $\theta_{mag}$
  & $\lambda_{obs}$ & $\phi_{orb}$ & Size 
  & Albedo\\
 & & &  & [$^{\circ}$]& [$^{\circ}$]& [$^{\circ}$]& [$^{\circ}$]& ['']
  & \%\\
\hline
7939 & 30 Oct 1998  & 1 & o53k01010 & 229.7 & 8.32 & 1.86 & 288.05 &
                                                                     1.71
  & 2.2 $\pm$0.4\\
 &  &  & o53k01020 &239.7 &7.39 & &288.67 \\
 &  & 2 & o53k01030 &276.3& 2.38 & &290.82\\
 &  &  & o53k01040&291.8 &-0.16 & &291.77\\
 &  & 3 & o53k01050&331.0&-6.13 & &294.22\\
 &  &  & o53k01060&345.6&-7.77 & &295.13\\
 &  & 4 & o53k01070&26.49&-9.45 & &297.62\\
 &  &  & o53k01080&39.52&-9.0 & &298.44\\
8224 & 23 Dec 2000  & 1&o5d602010&263.8 &4.31 & 3.09&102.99 & 1.75 &
                                                             1.9           $\pm$0.4\\
&   & &o5d602020&272.8 &2.93 & &103.54\\
&   & 2&o5d602030&308.6 &-2.91 & &105.63\\
&   & &o5d602040&323.2 &-5.09 & &106.53\\
9296 & 30 Nov 2003  & 1&o8m301010&275.3 &2.53 &-1.38 & 335.36 & 1.33 &
1.8    $\pm$0.5
  \\
&  & &o8m301020&285.1 &0.94 & & 335.95\\
&  & 2&o8m30103&322.5 &-4.99 & &338.15 \\
&  & &o8m301040&337.1 &-6.86 & &339.04\\
12244 & 19 Nov 2010  & 1&objy03010&174.3 &8.50 &2.12 &99.42 & 1.64 &
1.9    $\pm$0.3 \\
& & &objy03020&183.9 &9.09 & &100.19\\
& &2 &objy03030&218.6 &9.04 & &102.81\\
& & &objy03040&233.9 &7.96 & &103.99\\
& &3&objy03050&273.2 &2.87 & &106.94\\
& & &objy03060&288.5 &0.375 & &108.06\\
& 20 Nov 2010 &4&objy03070&327.8 &-5.72 & &110.86 \\
& & &objy03080&343.1 &-7.52 & &111.93\\
& &5&objy03090&22.45 &-9.5 & &114.61\\
& & &objy030a0&37.75 &-9.09 & &115.63\\
       & 01 Oct 2011  & 1&objy11010&164.8&7.69 & 3.6 & 89.51 & 1.78 &
                                                                      2.0 $\pm$0.3\\
& & &objy11020&174.5&8.52 & &90.10\\
& &2 &objy11030&210.7 &9.36 & &92.21\\
& & &objy11040&226.1 &8.59 & &93.15\\
& &3&objyb1010&272.7 &2.94 & &96.12\\
& & &objyb1020&282.4 &1.38 & &96.71\\
& &4&objyb1030&319.8 &-4.61 & &98.89\\
& & &objyb1040&335.3 &-6.66 & &99.84\\
& &5&objyb1050&14.89 &-9.45 & &102.26\\
& & &objyb1060&28.34 &-9.42 & &103.08\\
13328 & 23 Jan 2014  & 1&ocbug1010&145.2 &5.36 &1.77&78.90 & 1.7 &
1.5   $\pm$0.4 \\
&   & &ocbug1020&155.5 &6.69 &&79.54\\
&   & 2&ocbui1010&307.9 &-2.79 &&88.89\\
&   & &ocbui1020&318.2 &-4.37 &&88.53\\
& 27 Jan 2014  & 3&ocbug2010&10.39 &-9.34 &&270.22 &1.58 &
2.1 $\pm$0.5 \\
&   & &ocbug2020&20.75 &-9.5 &&270.86\\
& 25 Feb 2014  & 4&ocbuh3010&141.5 &4.85 &&275.98 & 1.7 &
2.0 $\pm$0.5 \\
&   & &ocbuh3020&151.8 &6.24 &151.8&276.62\\
14634 & 02 Feb 2017  & 1&od8k40010&197.2 &9.48 &-3.22&173.38 &1.45 &
1.7 $\pm$0.3\\
&   &2 &od8k40020&245.6 &6.75 &&175.92 \\
\hline
\hline
\end{tabular}
\end{table}
\change{ 
  Seven  
}{ 
  Six
} 
STIS campaigns were conducted during which Ganymede was observed
in the FUV range between 1150 {\AA} and 1700 {\AA}. All observations
were carried out with the G140L grating and used the Multi-Anode
Micro-channel Array (MAMA) detector. Due to Ganymede's synchronized
rotation around Jupiter, the various hemispheres are observable when
Ganymede is at distinct elongations on its orbit. Table
\ref{tab:exp_ovr} lists the available programs and Figure
\ref{fig:elong_ovr} shows the distribution of Ganymede's elongation
for the available datasets. For an impression of Ganymede's spatially varying
emission morphology we display in Figure \ref{fig:images}
selected observations at four
different orbital positions $\phi_{orb}$. They show Ganymede's leading,
trailing, sub-Jovian and anti-Jovian side. On the leading and trailing
side the ovals appear continuous across all visible longitudes, but on
the sub-Jovian and the anti-Jovian side, the emission appears
interrupted near 0$^{\circ}$ and 180$^{\circ}$ longitudes,
respectively. Observations near
180$^{\circ}$ have not been presented before in the literature to the
authors' knowledge. They give an impression that the auroral
brightness is not continuous along all longitudes of Ganymede, which
we will quantify further in Section \ref{s:results}.

\change{ 
  Fir  
}{ 
  For
} 
a complete map of the
auroral emissions, all datasets of the HST/STIS campaigns in table
\ref{tab:exp_ovr} were used to cover
all available elongations of Ganymede's orbit. 
 Ganymede was observed
on the downstream side around 90$^{\circ}$ elongation during 30
exposures and on the upstream side near 270$^{\circ}$ during 10
exposures.
\change{ 
  From the total 48 exposures available, the sub and anti-Jovian sides of  
Ganymede were observed only a small fraction of the time. Two exposures  
from the sub-Jovian side around 180$^{\circ}$ elongation were  
distorted and unusable due to a guide-star failure.  
}{ 
  Thus, of the 48 exposures, only 6 covered the sub- and anti-Jovian
  hemispheres. (Two additional exposures were distorted and unusable
  due to a guide-star failure.)
  } 
Therefore only six of the remaining 46 exposures covered the
regions around 0$^{\circ}$ and 180$^{\circ}$ elongation.

\subsection{Processing Auroral Disk Images}
The data analysis is performed with the flat-fielded detector counts from
the \texttt{.flt} files (see STIS instrument handbook, \citeA{rile17}). The major tasks to generate individual disk
images of Ganymede's auroral emissions include the determination of
Ganymede's position, size and orientation on the detector and
eliminating any solar-reflected and background emission photons which
are superimposed on the data.

Using the SPICE tool-kit along with additional information provided in
the scientific header of each file, we calculated the extension of
Ganymede's disk and its tilt on the detector, the system-III longitude and
magnetic latitude of Ganymede inside Jupiter's magnetic field, as well
as Ganymede's elongation around Jupiter to determine the hemisphere
observed in each exposure.

The size of Ganymede's disk on the detector varies between 53 to
nearly 80 pixels depending on Ganymede's distance to Earth. 
Ganymede's exact position within the 2 arcsecond slit (corresponding to 82
pixels) needs to be determined from the observations.
\change{ 
Therefore we carry out a Gaussian fit along the dispersion axis inside  
the  
Lyman-$\alpha$ window within which Ganymede's disk is represented by  
uniformly reflected solar photons.  
Similarly the central position  
along the slit is determined by a Gaussian fit along the spatial  
direction.  
} 
{ 
Therefore we use the Lyman-$\alpha$ emission on the detector, which primarily
consists of solar reflected light from Ganymede's surface and nearly spatially
homogenous emission from the geocorona. The position of Ganymede is
determined through a Gaussian fit along the direction of dispersion and along the direction of the slit.
} 
Due to a misalignment of the dispersion grating and the
detector the $y$ position of Ganymede $G_y (\lambda)$  is not 
constant along the dispersion axis. By performing Gaussian fits along
the spatial axis inside the Lyman-$\alpha$ window and the spectral trace
of the reflected solar light 
around 1600 - 1700 {\AA}, we calculate two different $y$ locations of
the disk that are used to estimate $G_y (\lambda)$ by linearly
interpolating between them.

To remove background emission in the form of dark pixels or
interplanetary noise we apply previously used techniques \cite{roth14,roth14a,saur15},
in which the average detector counts at each wavelength (i.e. each
pixel column)
\add{ 
  not affected by the signal from Ganymede
} 
are calculated and then subtracted from each pixel at
that column.
\remove{The background counts are averaged over 164 pixel from 
above and below the area that is associated with Ganymede's disk 
position. 
To guarantee that counts stemming from Ganymede do not influence the 
averaged background, a 10 pixel gap is added to the area of Ganymede's 
disk. For datasets where the disk was situated too close to the 
detector edge to allow for an averaging over 164 pixels, the maximum 
amount of pixel available to this edge were used.}
The solar reflected photons are removed by creating synthetic HST
datasets which contain these reflected solar photons, similar to
\citeA{musa17}. 
For each specific observation date the
measured solar spectra $f_s (\lambda)$ are retrieved from datasets of
the Upper Atmosphere Research Satellite (UARS) for observations older than
2001, and from the Solar Extreme Ultraviolet Experiment (SEE)
installed on the Thermosphere Ionosphere Mesosphere Energetic and
Dynamics orbiter (TIMED) for 2001 and later. Since the
retrieved spectra are measured at the Sun-Earth distance $d_{SE}$ they
are rescaled to resemble the photons reflected by Ganymede's disk
measured back at HST $f_{s,HST} (\lambda)$ by using
\begin{linenomath*}
\begin{equation}
f_{s,HST} (\lambda) = a \cdot f_{s} (\lambda) \cdot 
\left(
\frac{d_{SE}}{d_{SG} \, d_{GH}}
\right)^2 \, R_G^2 \qquad,
\end{equation}
\end{linenomath*}
where $d_{SG}$ and $d_{GH}$ are the Sun-Ganymede distance and
Ganymede-Hubble distance, respectively. From the reflected spectra, a
synthetic HST image is created
\change{ 
  that assumes Ganymede as a uniform 
reflecting disk across all wavelengths. Each photon flux $f_{s,HST} 
(\lambda)$ is redistributed into a spherical disk at the respective 
wavelength $\lambda$ and disk position along the slit with the 
appropriate disk size for each individual data set. 
}{
  by superposing photon flux $f_{s,HST}
  (\lambda)$ from uniformly reflecting disks for each wavelength.
}
\remove{
To incorporate the 
effects of light scattering and diffraction that is present on the 
real data, we convolve}
The
\add{ 
  resulting two-dimensional
} 
synthetic image is
then convolved with the point
spread function obtained by the \textit{TinyTim} software tool
\cite[]{kris11}. To match the unit of the synthetic data
$\phi_{refl}$, the measured detector counts $C_{obs}$ are converted to
photons cm$^{-2}$ s$^{-1}$ by dividing with the exposure time $t$ and
the effective HST primary mirror area $A_{HST}$ of 45,238 cm$^2$, as
given by
\begin{linenomath*}
\begin{equation}
\phi_{obs} = \frac{C_{obs}}{t} \cdot \frac{1}{A_{HST} \cdot T (\lambda)} \qquad.
\end{equation}
\end{linenomath*}
$T (\lambda)$ is the instrument dependent throughput that results in
the conversion from measured detector counts to effective photons
which reach the primary mirror. $\phi_{refl}$ and $\phi_{obs}$ are
then transformed into the one dimensional spectral flux densities
$s_{refl} (\lambda)$ and $s_{obs} (\lambda)$ in photons cm$^{-2}$
s$^{-1}$ \AA$^{-1}$ by the summation along the cross-dispersion axis
of the detector.
\remove{inside the pixels associated with Ganymede's 
  location.}
The geometric albedo $a$ is then calculated by performing a
least-square fit of $s_{refl}$ to $s_{meas}$ inside the wavelength
window of 1400 {\AA} to 1550 {\AA}, where all detected emission is
assumed to be due to solar reflected photons.
\add{ 
  The derived albedo values are listed in Table
} 
\ref{tab:exp_ovr}
\add{ 
  and are in agreement with those discussed in the
  literature for the different hemispheres by, e.g.,
} 
\citeA{feld00a,saur15,musa17,moly20}. 
 
\remove{The spectral image of reflected photons is scaled with the 
retrieved albedo to subtract the correct amount of solar photons from 
the observed data.}
The effective spectral image $\phi_{eff}$
displaying only auroral emission, i.e. that is free from background
emissions and solar reflected photons from Ganymede's surface is
calculated via
\begin{linenomath*}
\begin{equation}
\phi_{eff}  = \phi_{obs} - \phi_{back} - \phi_{refl} \qquad.
\end{equation}
\end{linenomath*} 
Before cropping the image to a 82 $\times$ 82 pixel sized array around
Ganymede's disk at 1356 {\AA} and rotating it to align with the
vertical axis, the image is converted into the unit Rayleigh (R)
which is defined as a surface brightness with 1 R = 10$^{6}$/4 $\pi$
photons cm$^{-2}$ sr$^{-1}$ s$^{-1}$, resulting in
\begin{linenomath*}
\begin{equation}
R = \frac{4 \pi}{10^6} \cdot \frac{\phi_{eff}}{m_s^2} \cdot
\left(
\frac{360 \cdot 3600}{2 \pi}
\right)^2 \qquad .
\end{equation}
\end{linenomath*} 
Here $m_s$ is the plate scale of the FUV-MAMA detector with the G140L
grating which is 0.0246
arcsec pixel$^{-1}$ \cite{rile17},
\add{ 
  and the last term represents
  conversion between arcsec and radian.
  } 

\subsection{Creating the Auroral Map}\label{sec:aur_map}
\begin{figure}[b!]
\noindent\includegraphics[width=\textwidth]{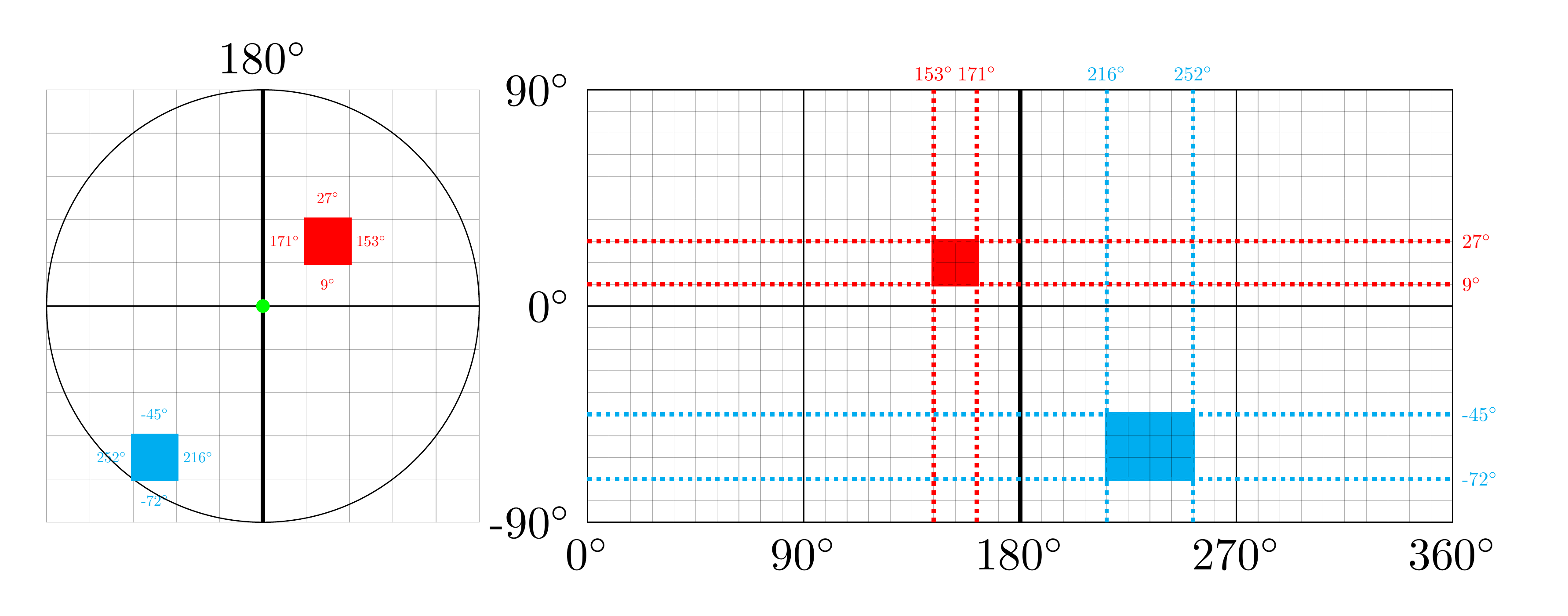}
\caption{Simplified illustration
  \add{ 
    of
  } 
  how pixels are mapped from
  the processed disk array (left) to a Mercator map (right). The
  examples shows a disk viewed at central meridian of 180$^\circ$ west
  longitude. 
After calculating the latitudes and longitudes of each pixel edge, the Rayleigh values are inserted into the pixel corresponding to the respective region.}
\label{fig:map_schem}
\end{figure}
From the final processed and rotated images a Mercator brightness map of the
aurora is created by mapping the brightness value of each pixel lying
on Ganymede's disk to a new position on an array which displays the
range of 0$^{\circ}$ - 360$^{\circ}$ west-longitude and -90$^{\circ}$
to  +90$^{\circ}$ planetocentric latitude of Ganymede. The map is created as a 360
$\times$ 720 pixel array, resulting in a resolution of 0.5$^{\circ}$
per pixel in both longitudinal and latitudinal directions. After
determining the area each pixel covers on Ganymede, the respective
Rayleigh value is mapped to the corresponding region on the projected map
as illustrated in Figure \ref{fig:map_schem}. The area that one pixel
covers is calculated from the latitude and longitude of its pixel
edges. 

To determine the longitudinal and latitudinal positions of all pixels on the disk, they are first mapped
on arcs of constant latitude. Because of the tilt of Ganymede as seen from the HST, the arcs are not
straight lines but sections of ellipses. Pixels of the same latitude therefore are not necessarily in the
same row. To infer the longitude of the pixel, the distance $d$ from the center, i.e. the sub-observer
longitude $\varphi_{sub-obs}$, along the arc is compared to the length of the whole arc $l$. The
longitude $\varphi$ can then be calculated as $\varphi = \varphi_{sub-obs} + \arcsin{2d/l}$. Note that
positions left from the center result in negative values of $d$, while positions right result in positive
values. Since the tilt of Ganymede is small, the arcs can be assumed to cover 180$^\circ$ in longitude.
Additionally we omitted sub-pixel discretization that
would account for the variation of latitude and longitude along the
pixel edges, but used the values of the pixel corners. 
Naturally this translates every disk
pixel to a rectified area on the Mercator map.

%
%

As the auroral emission is generated inside Ganymede's oxygen
atmosphere the brightness measured can be affected by the position of
a pixel on the disk. Due to the spherical
\change{ 
  extend
}{ 
  extent
} 
of the atmosphere,
photons registered by pixels near the edge of the disk can originate
from a larger atmospheric column than those of the central pixel below
the sub-observer point. To compensate for this effect, we scale the
brightness measured by each pixel with the atmospheric depth that lies
inside the line-of-sight of that pixel as observed from HST. The newly
shifted values then represent the brightness as observed from the
zenith of each location which eliminates the distortion obtained from
the viewpoint of HST. As 97 \% of Ganymede's oxygen atmosphere lies
below an altitude of $\sim$70 km \cite[]{marc07,saur15}, we use a length of
70 km for the sub-observer atmospheric height $H$ below the
zenith. Each disk pixel $R_{HST}$ is then adapted and shifted to the
zenith via
\begin{linenomath*}
\begin{equation}
R_Z (x,y) = R_{HST} (x,y) \cdot \frac{H}{L(x,y)} \qquad,
\end{equation}
\end{linenomath*} 
where $L$ is the atmospheric length of each pixel as viewed from HST.

After creating individual Mercator maps for each set of HST
observations,
all individual 46 maps were combined into one complete map. To create the final map that contains all exposures, the mean Rayleigh brightness for one pixel $x_m,y_m$ on the map is calculated from all Rayleigh values mapped to this pixel, weighted with the respective exposure time $T$ used for their observation as
\begin{linenomath*}
\begin{equation}
\bar{R}(x_m,y_m) = 
\frac{\sum_i^N \, R_i (x_m,y_m) \cdot T_i}
{\sum_i^N \, T_i}\qquad .
\end{equation}
\end{linenomath*} 
The averaged emission is  weighted with exposure time to generate the best temporal averaged
emission map in contrast to
\change{
  weighing 
}{
  weighting
}
with the inverse of the
uncertainty squared, which would correspond to
\change{
  weighing 
}{
  weighting
}
with intensity. 
$N$ is the number of exposures used for calculating the averaged brightness in
a pixel on the mercator map. Performing this for all 360
$\times$ 720 pixels on the map creates the final brightness map of
Ganymede's UV aurora at 1356 {\AA}.

From the Gaussian fits that were performed to locate Ganymede's disk
on the detector, an uncertainty of $\pm$ 1 pixel is estimated for the
deviation of Ganymede's central pixel. A deviation of one pixel could
already resemble a significantly different location assigned to a
pixel which is near the edge of the disk. We therefore only
incorporate disk pixel into the map whose assigned locations lie
within a defined window of uncertainty to assure a certain spatial
accuracy of the map. For that, the uncertainty in latitudinal and
longitudinal direction of each pixel is calculated for each
exposure. The uncertainties $\Delta x$ and $\Delta y$ describe the
total difference in latitude and longitude from both neighboring pixels
of the mapped cell. Due to the spherical curvature, $\Delta x$ and
$\Delta y$ are smaller at the disk's center and grow larger towards
the edge. We therefore chose a threshold value to filter pixel for
which the deviation of one cell would result in a larger spatial
discrepancy.  If any of the two uncertainties $\Delta x$ or $\Delta y$
exceeded a threshold of 15$^{\circ}$, the corresponding pixel is not
included into the map.

For additional evaluation tools, we use the same mapping procedure to
map the total exposure time that went into each pixel on the Mercator
map to assess
the observational coverage of different regions on the map. Similarly we create a
map for the signal-to-noise ratio (SNR) of each pixel on the Mercator map to
identify the data quality for later interpretations. The SNR is
calculated from the detector counts $C$, background emission counts
$B$ and solar reflected photons that are converted to detector counts
$S$. These components are mapped into an individual map as previously
described, but is unaffected by the atmospheric length correction and
exposure time weighting.
\change{
Unlike the mapped Rayleigh values, the total 
counts contained in one pixel of the disk are evenly re-distributed 
over all corresponding pixels on the map.
}
{
  Since the total number of counts needs to be conserved, the counts
  are evenly re-distributed over all corresponding pixels on the
  map. This is contrary to the mapping of the Rayleigh values, where
  the average brightness over all corresponding pixels is considered.
}
The three individual maps
for $C$, $B$ and $S$ are then combined via
\begin{linenomath*}
\begin{equation}
SNR = \frac{C - B -S}{\sqrt{C + B + S}} \qquad ,
\end{equation}
\end{linenomath*} 
to create a complete SNR map.

\section{Results}
\label{s:results}
\begin{figure}[b!]
  \noindent\includegraphics[width=\textwidth]{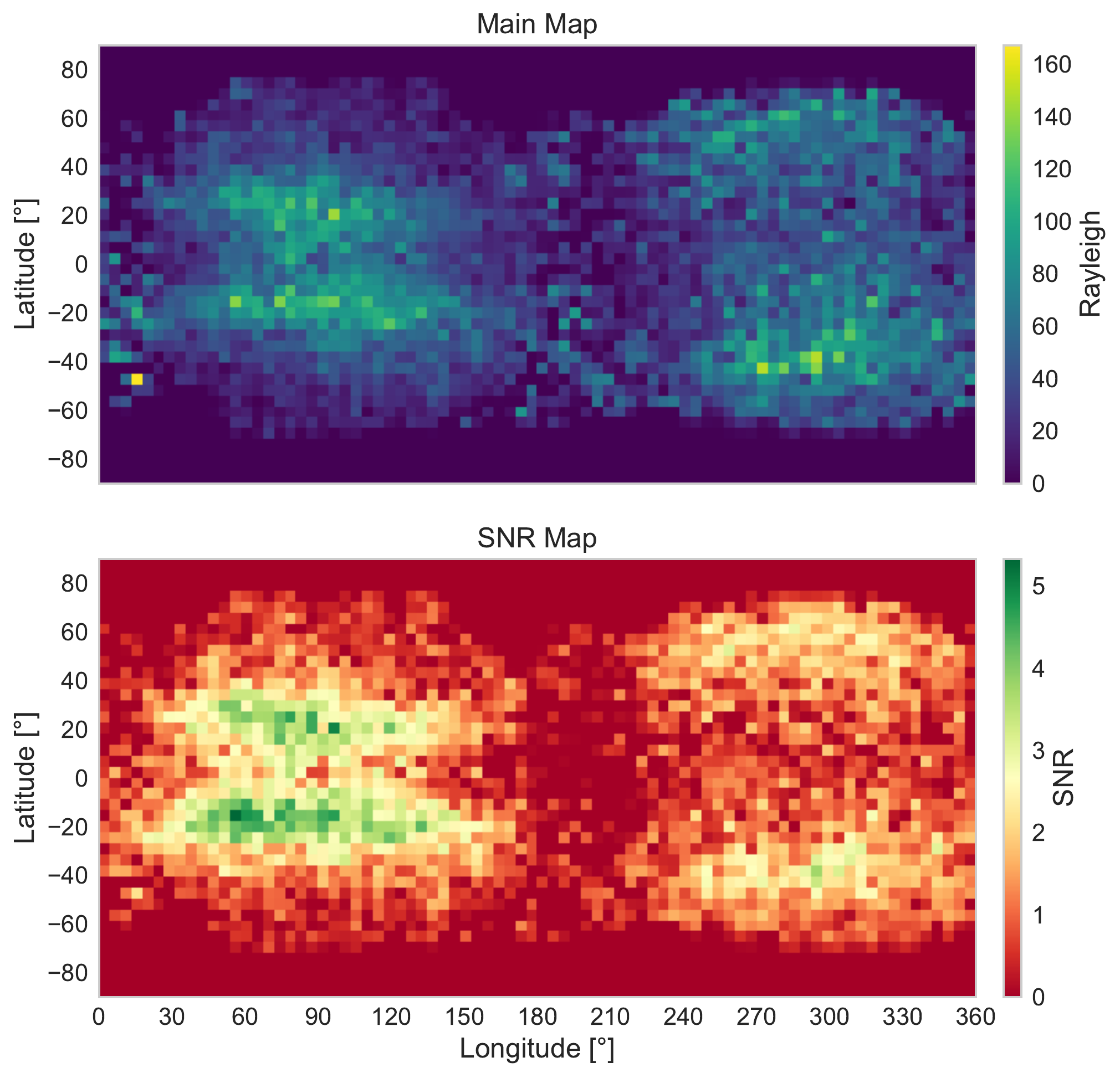}
\caption{Main brightness map at 1356 {\AA} that incorporates all
  \remove{used} 
  46 exposures from Table \ref{tab:exp_ovr} (top) and the corresponding signal-to-noise map (bottom).}
\label{fig:main_map}
\end{figure}
In this Section we  present our calculated brightness map of
Ganymede's auroral emission at 1356 {\AA}. We also analyze its
properties and discuss the possible mechanisms responsible for its
spatial structure.

\subsection{Main brightness map}
In Figure \ref{fig:main_map} we display the main, i.e. averaged
brightness map of Ganymede's aurora. 
The map was rebinned to cells which
contain 9 $\times$ 9 pixel of the unbinned map in order to increase
the SNR. The rebinned map
therefore has a resolution of 40 $\times$ 80 pixel, where one pixel
spans 4.5$^{\circ}$ $\times$ 4.5$^{\circ}$ in latitude and
longitude. From simple visual inspection of the brightness map, the
auroral emission seems to be clearly dominant on the 
downstream and upstream hemispheres, while the transition regions
appear noticeably fainter. The SNR map displayed in the bottom part of
Figure \ref{fig:main_map} also represents this aspect to some extent as the
signal-to-noise ratios are clearly higher on the upstream- and
downstream sides, compared to the sub- and anti-Jovian longitudes
around 0$^{\circ}$ and 180$^{\circ}$, respectively. Note however that the SNR is
large when the photon fluxes and/or the exposure times are large.
With a total exposure of $\sim$
5,000 to 7,000 seconds for the sub- and anti-Jovian sides, the low SNR
of $\leq$1 of individual pixels on the map 
indicate that very few photons could be detected at these
longitudes.

\subsection{Brightness maps: Inside, above and below the current sheet}

\begin{figure}
\noindent\includegraphics[width=\textwidth]{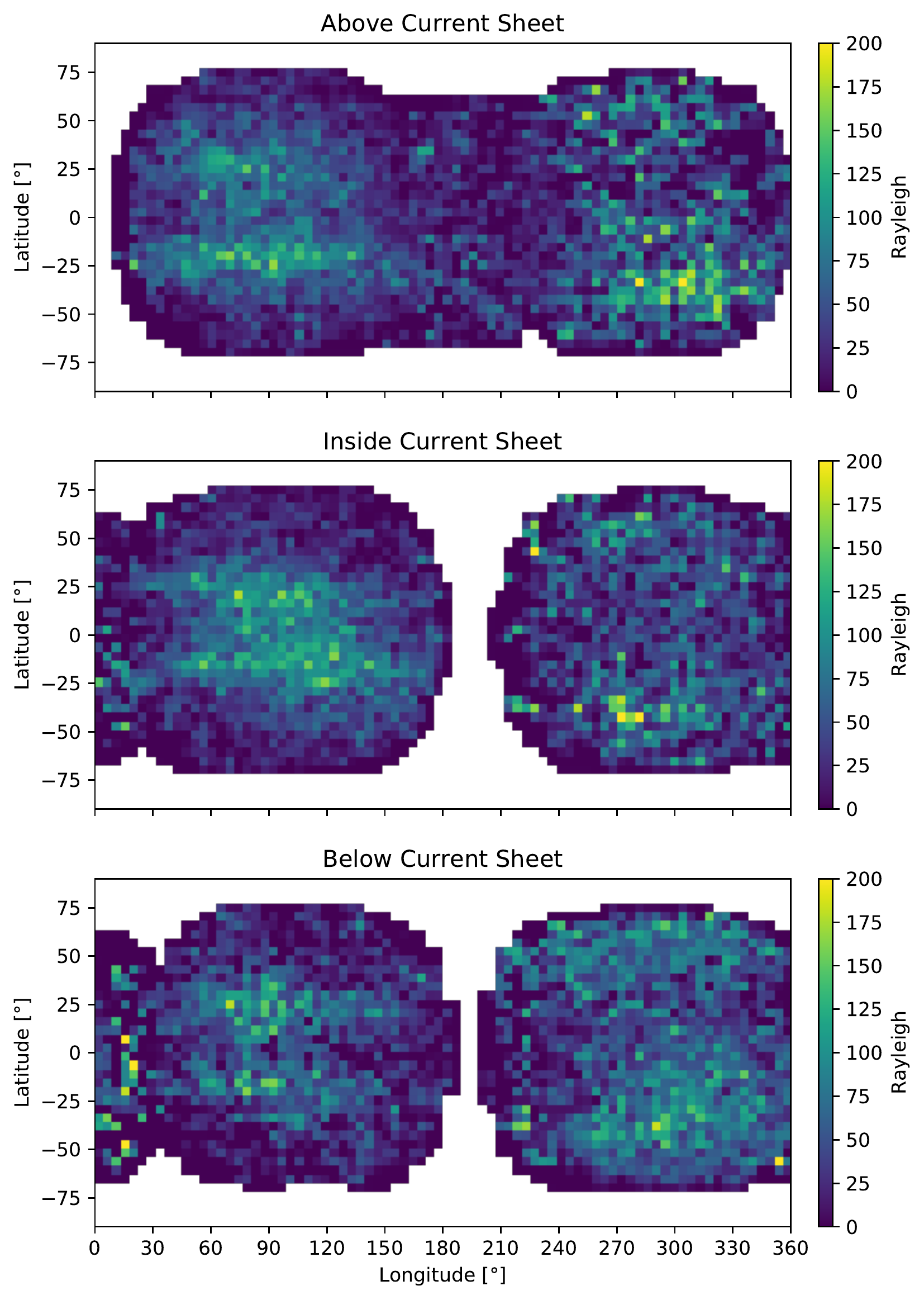}
\caption{Comparison of the evaluated 46 exposures separated by
  Ganymede's magnetic latitude when it is above ($\theta_{mag} >
  6^{\circ}$), inside ($|\theta_{mag}| \le 6^{\circ}$)  or below ($\theta_{mag} < -6^{\circ}$) the Jovian current sheet.}
\label{fig:io_comp}
\end{figure}
\begin{figure}
\noindent\includegraphics[width=\textwidth]{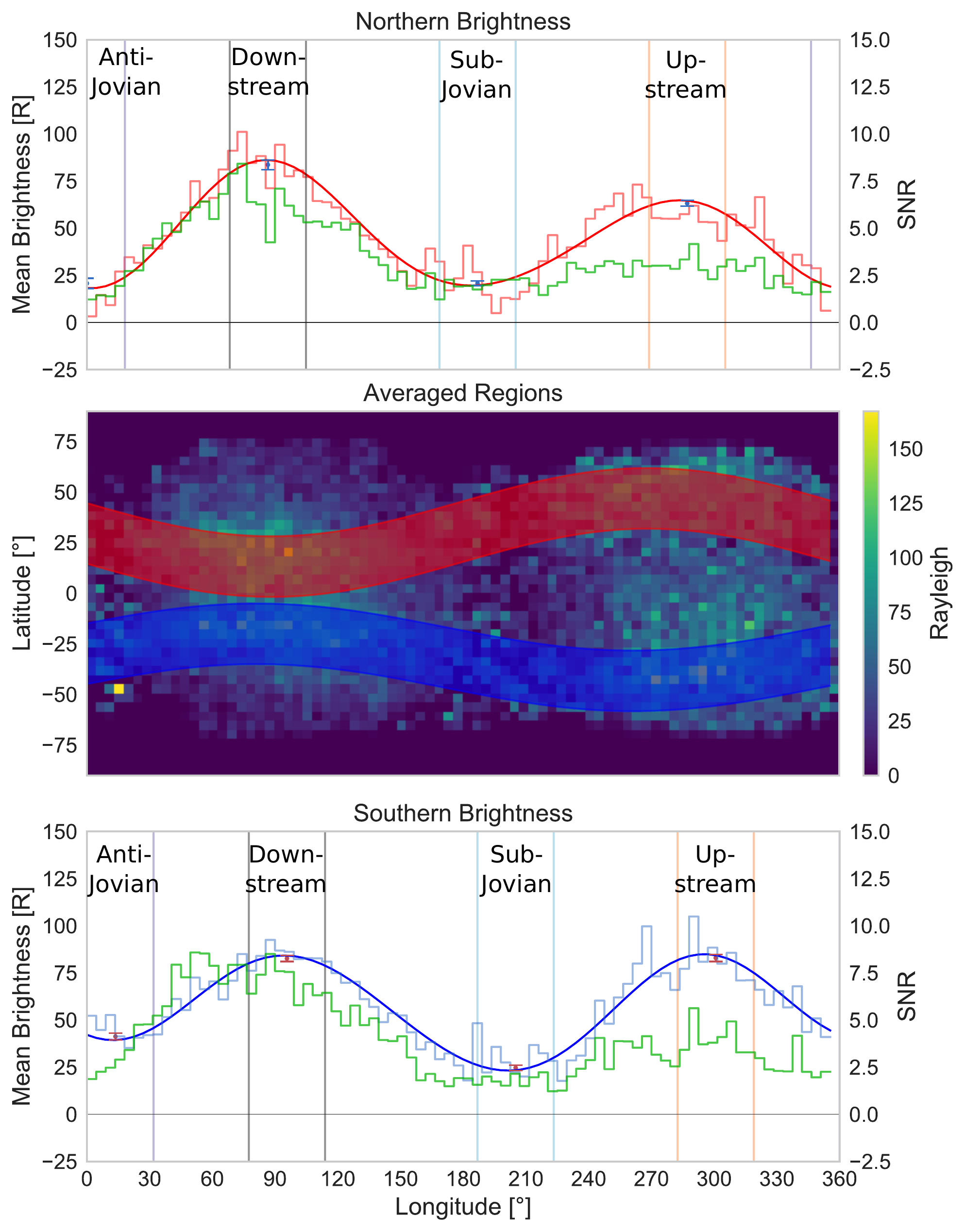}
\caption{Average auroral brightness (
  \add{ 
    based on all available exposures)
  } 
  for the northern (top, red) and
  southern (bottom, blue) ovals as a function of longitude. The 
average brightness is approximated by sinusoidal based fit functions
(also red and blue). The signal-to-noise ratio
for the averaged regions is plotted in green.
\remove{
The center 
  panel highlights the regions (red and blue bands) 
  used for averaging within the original brightness map from Figure}
\add{ 
  The center panel is a replot of the main brightness map (Figure
} 
\ref{fig:main_map})
\add{ 
  and includes as red and blue bands the oval regions used to
  calculate the values in the top and bottom panel (more details see
  Section
} 
\ref{ss:ana}). Regions with a width of 36$^{\circ}$ longitude used to
  calculate average peak and faintest emission are indicated by
  vertical bars in the top and bottom panels, and the points with
  error bars indicate average values in these longitude ranges (Table
  \ref{tab:final_comp}).
}
\label{fig:bband}
\end{figure}

Figure \ref{fig:io_comp} displays the brightness maps which were
created when we separated the available exposures according to the
magnetic latitude of Ganymede in Jupiter's magnetosphere. 
We choose as boundaries for the current sheet $\theta_{mag} = \pm
6^{\circ}$ magnetic latitude. 13 exposures make up the map below the current
sheet, 14 above the current sheet, and the remaining 19 exposures are
used for the map where Ganymede is located within the current
sheet. The maps show that the coverage of the main brightness map in Figure
\ref{fig:main_map} is not evenly
distributed for all magnetic latitudes. The longitudinal region
between 180$^{\circ}$ and 210$^{\circ}$ is only covered by
observations where Ganymede is above the current sheet. The comparison
illustrates the prominent emission structures on the downstream and
upstream sides, as well as the enhanced upstream emission on the
southern oval across all three maps. Increased values around the
0$^{\circ}$ sub-Jovian longitude are only present in isolated pixels
when Ganymede is below the current sheet and are not
visible on the other two maps. They could be either a
non-systematic, sporadic effect or an unknown systematic difference
between northern and southern latitudes.  However,  the SNR map in
Figure \ref{fig:main_map}
shows that the fluxes of these pixels are barely significant due to
the low exposure times and thus no firm conclusions can be reached. 

\subsection{Analysis of the brightness map}
\label{ss:ana}

A global fit for the
latitudes $\Theta $ of the ovals as a function of all longitude $\lambda$
incorporating all exposures is generated in the form 
\begin{linenomath*}
\begin{equation}
\Theta(\lambda) = \Theta_0 + \Theta_1 \sin(\lambda + \lambda_1), 
\label{e:Theta}
\end{equation}
\end{linenomath*}
where $\lambda$ is the western longitude and $\Theta_0, \Theta_1$ and
$\lambda_1$ 
are the fitting parameter summarized in Table \ref{t:fit}. 
%
\begin{table}
 \caption{Fit values for expressions (\ref{e:Theta}) and (\ref{e:B}). }
 \centering
 \begin{tabular}{l c c c c c c c c}
 \hline
  Eq  & hemisphere &  $\Theta_0$ & $\Theta_1$  & $\lambda_1$  \\
 \hline
  (\ref{e:Theta}) & north &    32.3$^{\circ}$ &   -16.9$^{\circ}$ &  1.5$^{\circ}$  \\
   (\ref{e:Theta})  & south &  -29.4$^{\circ}$ & 11.6$^{\circ}$ &
                                                                  8.8$^{\circ}$ \\
\hline
    \\
\hline
  Eq  & hemisphere &  $B_0$ & $B_1$ & $B_2$& $B_3$ & $\lambda_1$  & $\lambda_2$& $\lambda_3$\\
 \hline
(\ref{e:B}) & north   & 46.2 R &  9.4   R &  28.2   R & 5.2 R & 19.9$^{\circ}$ & 264.5$^{\circ}$& 236.1$^{\circ}$\\
(\ref{e:B}) &  south &  57.0 R & 11.0 R &  25.8   R & 3.6 R  & 65.7$^{\circ}$& 237.4$^{\circ}$& 256.9$^{\circ}$\\
 \hline
 \end{tabular}
\label{t:fit}
 \end{table}
Average latitude values for the ovals are calculated
inside the same longitude windows of
40$^{\circ}$ to 150$^{\circ}$ and 240$^{\circ}$ to 340$^{\circ}$ for
both southern and northern emission.
\add{ 
  The windows are slightly off centered from 90$^{\circ}$ and
  270$^{\circ}$ due to the shifted minima (see Figure
  } 
\ref{fig:bband}).
The average latitudes
on the downstream hemisphere of 
{$\pm$18.7$^{\circ}
  \pm 4.5^{\circ}$}
 as well as the mean upstream latitude of
{ $\pm$$41.5^{\circ} \pm 6.7^{\circ}$} 
are in 
accordance with the reported locations in 
\citeA{mcgr13}. 

To further compare our results with previous works, we first study the average positions
of the northern and southern ovals when Ganymede is  inside the current
sheet and outside of it. Therefore we fit
polynomials 
of second degree within downstream longitudes 
of  40$^{\circ}$ to 150$^{\circ}$ and upstream longitudes of  240$^{\circ}$ to 340$^{\circ}$
using a
centroiding scheme as in \citeA{saur15}. 
The averaged latitudes are
shifted by 
 $-5.4^{\circ} \pm 3.2^{\circ}$
towards the equator on the downstream hemisphere when Ganymede
is located inside the current sheet compared to outside. 
The retrieved shift of $\pm 5.4^{\circ}$  is in reasonable agreement with the
shift of $-4.1^{\circ} \pm 0.7^{\circ}$  found in \citeA{musa17}. 
The detected shift by
\citeA{musa17} towards the poles on the upstream hemisphere is
reproduced in our results only on the southern oval, and is not
clearly observable on the northern oval location.

\change{
  For the average disk brightness of all exposures, 
}
{
  Combining all exposures, we calculate a disk averaged brightness and
  find
}
values of $68.3 \,\pm\, 8.9$ R on the upstream and $90.5 \, \pm\, 6.4$
R on the downstream side. 
Comparing the auroral brightness from our maps when
Ganymede is inside or outside of the current sheet, we calculate that
the brightness on the downstream aurora increases by a factor of
$1.3\, \pm\, 0.31$ as Ganymede transitions into the current sheet and
decreases by a factor of $0.78\, \pm\, 0.19$ on the upstream
side. Both values are well in agreement with the results of
$1.33\, \pm\, 0.05$ and $0.76\, \pm\, 0.07$  by \citeA{musa17}.

To further characterize the emission structure of the northern and
southern auroral emission, 
we fit the brightness
\change{
  along the bands 
}{
  within the bands displayed in Figure
}
  \ref{fig:bband}
in the form
\begin{linenomath*}
\begin{equation}
B(\lambda) = B_0 + B_1 \sin(\lambda + \lambda_1) + B_2 \sin(2 \lambda + \lambda_2) + B_3 \sin(3 \lambda + \lambda_3)
\label{e:B}
\end{equation}
\end{linenomath*} 
 with the fitting parameters $B_0, B_1, B_2, B_3,\lambda_1,\lambda_2$ and
 $\lambda_3$ provided in Table \ref{t:fit}. The fit is based on the
main map (Fig. \ref{fig:main_map}), where we used
the brightness of bins at position
 $\Theta(\lambda)$  from expression (\ref{e:Theta}) plus
 its three latitudinally neighboring bins above and below. The latitudinal
 extension corresponds to approximately  31$^{\circ}$ and the
 associated  band is highlighted on the map in Figure
 \ref{fig:bband}. With expression (\ref{e:B}), we introduce a fit
 function with 7 free parameters in order to resolve
various asymmetries in the brightness distributions.
The observed and fitted brightnesses are displayed in the top and bottom panels of Figure
\ref{fig:bband}  along with the integrated SNR of those regions in green. For both the northern
and southern ovals, the averaged brightness exhibits a 
sinusoidal shape without abrupt drops or cut-offs, which can also be
observed in the SNR. From the brightest peaks on the
down- and upstream sides, the brightness steadily decreases 
towards  $\sim$0$^{\circ}$ and $\sim$180$^{\circ}$ longitudes
regions where they
reach their lowest values. 

In order to quantify the brightness change along the ovals we 
average the brightness inside windows of 36$^{\circ}$ longitude around
the fitted brightest and faintest points along the sinusoidal fits.
\add{ 
  The widths of these windows were chosen such that enough data points
  lead to a robust value and that the widths are still narrow enough
  such that the minimum and maximum are approximated well.
} 
The uncertainties for those values is calculated from the variance of the
brightness inside those 36$^{\circ}$ windows.
\begin{table}[b!]
	\caption[Overview of brightness variations]{ Averaged
          brightness within various longitudinal ranges and their ratios. Brightness is given in
          units of Rayleigh (R). See text for details on averaging.
          \add{ 
            Downstream/upstream and
          sub-Jovian/anti-Jovian averages are referred to as joint
          brightnesses, respectively. They are provided
          together with  the north-south averages for a basic overview.
        } 
    }    \label{tab:final_comp}
	\resizebox{\columnwidth}{!}{%
	\begingroup
\setlength{\tabcolsep}{10pt} 
\renewcommand{\arraystretch}{1.5} 
	\begin{tabular}{c c |c c |c c |c }
	\hline
	\hline
	 &  & \multicolumn{2}{c|}{downstream /  upstream}
          &\multicolumn{2}{c|}{Jovian side / anti-Jovian side} & Ratios\\
	\hline
	\hline
	\multirow{3}{*}{Northern} & Longitude Range & 68$^{\circ}$ --
                                                      105$^{\circ}$&
                                                                     269$^{\circ}$
                                                                     --
                                                    305$^{\circ}$&
                                                                   346$^{\circ}$
                                                                   --
                                                                   18$^{\circ}$&
                                                                                 168$^{\circ}$
                                                                                 --
          205$^{\circ}$\\
	 & Brightness within Range   & $ 83.8 \pm 2.6$ R& $63.5 \pm
                                                         1.6$ R & $19.6 \pm 2.8$ R& $20.3 \pm 1.3$ R\\
	  & Joint Brightness  & \multicolumn{2}{c|}{$73.7 \pm
                                       1.5$ R} &
                                                 \multicolumn{2}{c|}{$19.9
                                                 \pm 1.5$ R} & $3.7 \pm 0.4$ \\
	 \hline 
	 \multirow{3}{*}{Southern} & Longitude Range& 77$^{\circ}$ --
                                                   114$^{\circ}$&
                                                                  283$^{\circ}$
                                                                  --
                                                                  319$^{\circ}$&
                                                                                 0$^{\circ}$
                                                                                 --
                                  32$^{\circ}$& 187$^{\circ}$ -- 223$^{\circ}$\\
	 & Brightness within Range & $82.7 \pm 1.6$ & $83.0 \pm 1.9$
                                                       R& $40.9 \pm
                                                          1.8$ & $24.5
                                                                 \pm
                                                                 1.4$ R\\
	 & Joint Brightness & \multicolumn{2}{c|}{$82.9 \pm 1.2$ R} &
                                                                      \multicolumn{2}{c|}{$32.7
                                                                      \pm
                                                                      1.1$
                                                                      R} & $2.5 \pm 0.2$ \\
	 \hline
	 \multicolumn{2}{c|}{North-South Average} &
                                                    \multicolumn{2}{c|}{$78.3
                                                    \pm 1.0$ R} &
                                                                  \multicolumn{2}{c|}{$26.3
                                                                  \pm
                                                                  0.9$
                                                                  R} & $3.0 \pm 0.1$ \\
	 \hline
	 \hline
	\end{tabular}
	\endgroup	
	}
\end{table}
%
%
Average brightness values within various longitudinal regions are
quantitatively provided in Table \ref{tab:final_comp}. The values are
calculated as algebraic averages within a band given by 
the bin with the maximum brightness 
$\pm3$ bin in latitudinal direction
and within the longitude ranges specified
in the table. The area for each region is approximately
36$^\circ$  $\times $31$^\circ$. 
Within the northern oval, the emission decreases from 83.8 $\pm$2.6 R
on the downstream side and 
63.5 $\pm$1.6 R on the upstream side
to small values
 of $19.6 \, \pm \,
1.3$ R and
20.3 $\pm$ 2.8 R on the sub- and anti-Jovian longitudes, respectively.
The emission on  the flanks (i.e., Jovian and anti-Jovian sides) is therefore
a factor of 3.27$\pm$ 0.4 fainter than on the
up- and downstream sides. 
For the southern oval, the upstream and downstream brightness is
similar  with an
average value of 82.9 $\pm$ 1.2 R. 
The average faint emission around 0$^{\circ}$ and
180$^{\circ}$ is significantly stronger on the southern oval with an
averaged brightness of 32.7 $\pm$ 1.1 R compared to the northern hemisphere. 
While the decrease towards the
sub-Jovian hemisphere is only a factor of $2.0\, \pm\, 0.1$, where the emission
is still 40.9 $\pm$ 1.8 R, the brightness decreases by a factor of
3.5 $\pm$0.2 towards the anti-Jovian longitude where the auroral brightness is only 24.5
$\pm$ 1.4 R. For the southern oval, we find an average brightness change by a factor of  2.5 $\pm$ 0.2 when
comparing the averages of the trailing and leading sides to the
flanks. Finally, combining the emission from the northern and the southern ovals within the
individual longitudes given in Table \ref{tab:final_comp}, we find
the emission on the flanks is a factor of 3.0 $\pm$0.1 lower
compared to the average oval brightness of the upstream and downstream side.

The main map in Figure \ref{fig:main_map} shows that the maximum
brightness is not exactly located at 90$^\circ$ and 270$^\circ$ longitudes,
i.e., symmetric with respect to the Jovian and anti-Jovian side. On
the downstream side the maximum is at 85$^\circ$ for the northern band
and at  95$^\circ$ for the southern band. On the upstream side the
emission maxima lie at 283$^\circ$ for the northern band and at
297$^\circ$ 
for the southern band, i.e. maximum brightness is shifted
towards the Jovian-facing side by 20$^\circ$ on average. The reason for this asymmetry could
lie in the slightly tilted magnetic moment of Ganymede \cite{kive02}
and/or in asymmetries of the plasma interaction, e.g., due to the Hall
effect \cite{dore15,saur99a}.

On the anti-Jovian flank, brighter regions appear to be present around
longitude 190$^{\circ}$, in both the northern and the southern regions, embedded inside the faint 
aurora. They   also appear inside the brightness curves in the top and bottom
panels of Figure \ref{fig:bband}. The peaks are not correlated with a similar increase in the
signal-to-noise ratio due to low total exposure times covering this
region. Therefore it is doubtful if the locally enhanced
brightness patches are physically real.


For several reasons, the largest values within our auroral brightness map are
smaller than  
previously reported values in the range of 100 R up to 300 R  in locally bright
areas in \citeA{feld00a}. For one we retrieved the values from our
auroral map instead of the observed disks. Since the map
incorporates multiple exposures into a weighted average of each pixel
on the map, any individual high-count emission from a detector pixel of one exposure
gets averaged by exposures which went into the same map pixel with
fewer detected counts. Additionally, unlike studies where the observed
disks were evaluated, we accounted for the atmospheric line-of-sight
effect described in Section \ref{sec:aur_map} when creating the
map. Thus high brightness pixels near the edges of the disk are given
a lower adapted-brightness on our map. Lastly the rebinning of our map
to increase the SNR value affects the brightness as it averages
individual bright spots. Since the actual size chosen to rebin has a
direct impact on the brightness averaging, our size of $3 \times 3$
pixel used for rebinning  exceeds the rebinning size of $2 \times 2$
used \citeA{mcgr13} on the disks.

\subsection{Interpretation of the auroral brightness map}
\begin{figure}
\noindent\includegraphics[width=13cm]{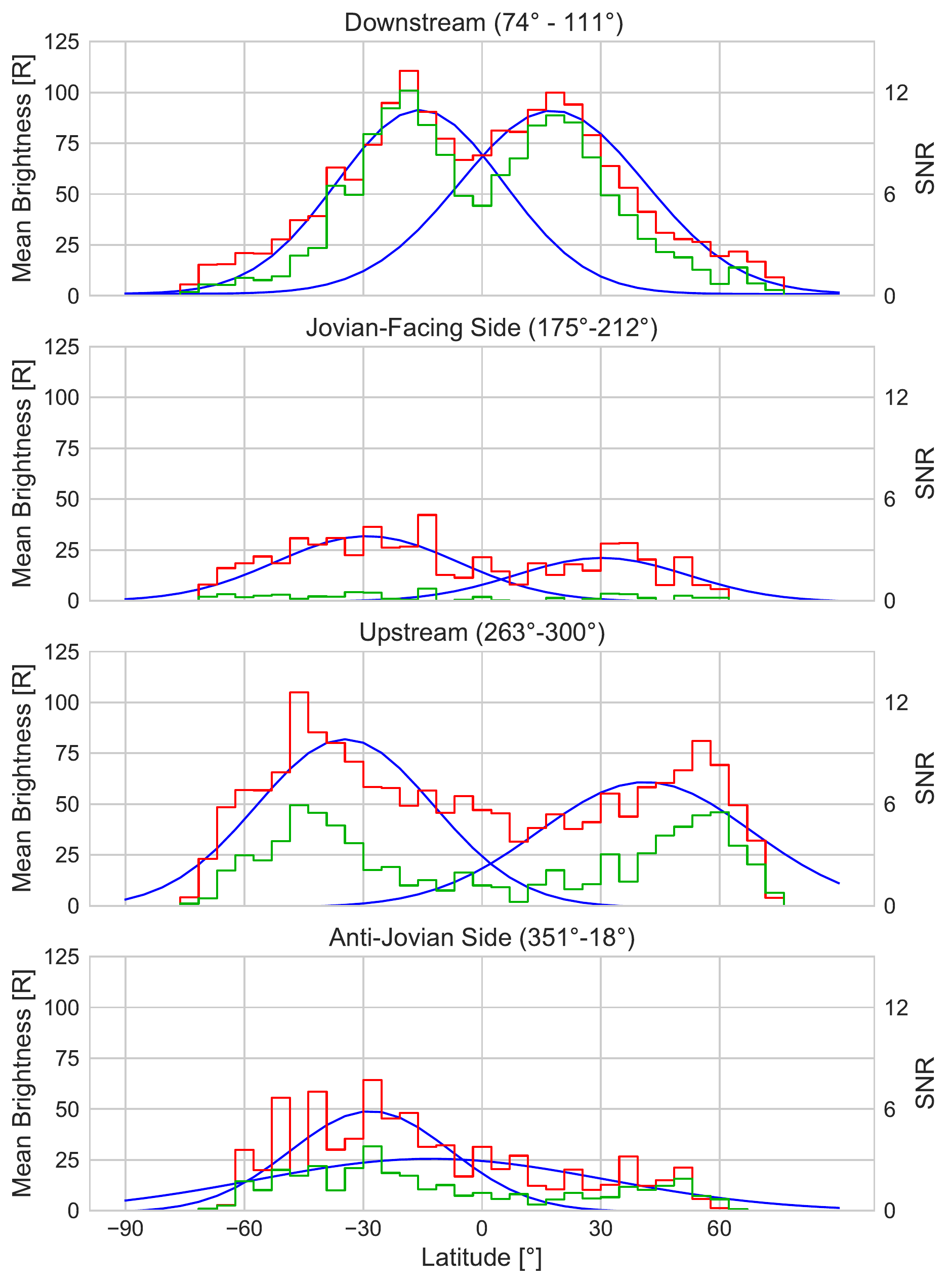}
\caption{Mean brightness as function of latitude for the downstream,
  Jovian-facing side, the upstream side and the anti-Jovian side (in
  red). For the mean brightness, latitudinal bands within a width of
  three bins 
have been used.  The curves in green display the
  associated SNR.  In blue a Gaussian fit to the mean brightness
  within the northern and southern hemispheres is overlaid. 
}
\label{fig:main_fits}
\end{figure}
There are different hypotheses on the cause of Ganymede's aurora
and therefore there are also different ways to
interpret the derived brightness maps of Ganymede's aurora in Figure
\ref{fig:main_map}. To point out
different possible interpretations, we display the UV brightness as a function of
latitude  in Figure \ref{fig:main_fits} for the upstream, the
downstream and the sub- and anti-Jovian sides, respectively. For the
upstream and downstream sides a similar analysis has been performed by
\citeA{musa17}. On the upstream and downstream sides the brightness has two maxima, respectively,
consistent with the existence of two auroral bands in the north and
south.  We display these
structures by separately fitting Gaussians on the northern and southern hemisphere,
respectively (shown as blue lines in Figure \ref{fig:main_fits}).
The brightness maxima are closer
to the equator on the downstream side compared to the upstream side 
due to the magnetic stresses of Jupiter's magnetospheric plasma on
Ganymede's magnetospheric plasma. The downstream aurora is also brighter
on average compared to the upstream side. 

The brightness distribution as a function of latitude 
on the Jupiter facing side and the anti-Jupiter
side shown in Figure \ref{fig:main_fits} is less clearly structured. 
The anti-Jupiter side has two modest maxima in the
northern and southern hemispheres with slightly reduced values around
the equator. The Jupiter facing side shows two modest
maxima in the south and not in the north, while for the anti-Jupiter
side a maximum is only visible in the south.
\add{ 
  Emission from latitudes larger than $\pm$70$^\circ$ latitudes are
 difficult to assess due to the viewing geometry
 from Earth.
} 

Auroral emission displayed in  Figures \ref{fig:main_map} and
\ref{fig:main_fits} maximizes within
upstream and downstream northern and southern crescents away from the
equator. But
\remove{spatially 
  varying}
auroral emission
with a smaller amplitude is still present within all  other 
\add{ 
  longitudes.
  } 
\remove{regions within $\pm$70$^\circ$ latitude. 
The amount of emission from 
higher latitudes is difficult to assess due to the viewing geometry 
from Earth. }
Several possible scenarios causing the auroral emission pattern
are possible. 
\begin{enumerate}
\item One possibility is that reconnection near the open-closed field line boundary
generates energized particles which propagate
along the  field lines and generates northern 
and southern auroral crescents on the upstream and
downstream side, respectively. 
Maximum reconnection is expected to occur based on
numerical simulations and theory on the upstream and downstream side
\cite[]{neub98, duli14, toth16, zhou19}. Consequently the
reconnection intensity gets weaker towards the flanks of the sub- and
anti-Jovian hemisphere, where the plasma flow is parallel to the magnetopause
 and the exerted ram pressure diminishes.
Energetic particles will however drift
perpendicular to the field lines and might be scattered and thus
additionally diffuse
across the field lines to generate auroral emission on field lines
located away from the reconnection sites.  This could be a scenario
explaining the
\change{
  none-negligible 
}{
  non-negligible,
  }
but
weak emission on the flanks compared to the upstream and downstream
side and the weak emission near equatorial latitudes.
\item 
Alternatively, several different auroral generator mechanisms could contribute to Ganymede's 
auroral emission.  Next to reconnection on the upstream and downstream
side, shear flow near the open-closed field boundary could
drive an electric current system with field-aligned electric current
predominately towards Ganymede's ionosphere on the flanks
\cite<e.g.,>[]{jia09}. 
These currents might drive
parallel electric fields which accelerate particles subsequently
creating the aurora \cite[]{evia01}. The existence and nature of
such DC parallel electric fields similar to
observations and theory from Earth
\cite{knig73} is however uncertain at Ganymede. Within the
closed-field region of Ganymede's magnetosphere, 
possible MHD and plasma waves could be subject to wave-particle
interaction and thus produce energetic particles  
\cite<e.g.,>[]{evia01, lysa96, saur18a}.
Additionally, on open
field lines, energetic ions and electrons from Jupiter's magnetosphere
will contribute to Ganymede's polar cap auroral emission. Several of
these processes thus could jointly shape Ganymede's auroral structure.
\item
The local auroral emission rate  also depends on the neutral
density. The primary component of Ganymede's atmosphere is O$_2$ with
a contribution from H$_2$O near the sub-solar point
\cite{hall98,marc07,roth21}. The spatial variability and composition of the
atmosphere has been modeled by, e.g.,
\citeA{coll18,lebl17,carn19,plai20}.
\change{
  The primary component of Ganymede's 
}{
  The
}
atmosphere's O$_2$ is however expected  to only
weakly vary across the surface of Ganymede because O$_2$ does not
freeze out on the surface \cite<e.g.,>[]{stro05}. The spatial variability
of the other neutral components is thus expected to contribute to the
spatial variability of Ganymede's UV emission.
\end{enumerate}


\section{Summary}
In this work we used a set of 46 exposures taken with the STIS
instrument of the Hubble Space Telescope from 1998 to 2017 to create a
global brightness map of Ganymede's auroral emission at
1356{\AA}. Our results are consistent with the location map of
\citeA{mcgr13} and the brightness values derived in
\citeA{musa17}. The map and analysis of this work shows that the
brightness of Ganymede's aurora varies strongly with 
longitude. With strongest emission on the upstream and downstream
sides around 90$^{\circ}$ and 270$^{\circ}$ longitude, the emission
around the sub- and
anti-Jovian longitudes near 0$^{\circ}$ and 180$^{\circ}$ are on
average 3.0 times 
fainter. While the
brightness does not completely vanish, thus making the aurora not
strictly discontinuous, the northern and southern emission can each
be characterized to consist of two dominant auroral crescents rather than a
continuous oval. Compared to other celestial bodies in our solar
system which exhibit auroral emission like Earth, Jupiter, Saturn
and Uranus 
\cite<e.g.,>{bhar00, clar05, lamy12}
the distinctively cresent-shaped contributions to its
auroral ovals makes Ganymede aurora unique in the solar system

This study presents new observational constraints
on Ganymede's auroral ovals. 
%
%
%
The derived auroral maps are maps of Ganymede's
magnetospheric physics, which  will be helpful for future
investigations of Ganymede's mini-magnetosphere and its auroral
acceleration processes.
\add{ 
  For example, it will be interesting to relate the spatial
  distribution of the auroral
  emission to the in-situ magnetic field and plasma measurements by
  the Galileo spacecraft
} 
\cite<e.g.,>{kive02,evia00,coll18}.
They will be useful for a comprehensive
understanding of Ganymede  and for the planing of future measurements
taken by the JUICE spacecraft \cite{gras13} and for interpretation of
observation by the Juno spacecraft \cite{bolt17}. These observations
will help to provide an in depth understanding
of Ganymede's magnetosphere and internal structure, but also its
coupling to Jupiter \cite{bonf17} and its  influence of Jupiter's
magnetosphere.
The sub-Alfv\'enic aurora of Ganymede - the only sub-Alfv\'enic one in the solar
system - might also be a model case for sub-Alfv\'enic aurora on close-in
exoplanets \cite<e.g.,>{zark07,saur13,saur21}.

\section{Open Research}
 All data used in this study is available on the Mikulski Archive for
Space Telescopes (MAST) of the Space Telescope Science Institute at
\url{http://archive.stsci.edu/hst/}.  The specific datasets
\add{ 
  used here
} 
are listed in
Table \ref{tab:exp_ovr} and can be accessed at:
\citeA{moos97}, 
\citeA{mcgr99},
\citeA{ford02},
\citeA{saur10a},
\citeA{nich13},
and \citeA{grod16}.

The data for Figures 5 to 8 can be accessed at \citeA{marz22}.

\acknowledgments
\noindent This project has received funding from the European Research
Council (ERC) under the European Union’s Horizon 2020 research and
innovation programme (grant agreement No. 884711).

%
%


\end{document}